\documentclass{aa}

\usepackage{newtxtext,newtxmath}
\usepackage[T1]{fontenc}
\usepackage[dvipsnames]{xcolor}
\usepackage{wasysym}
\usepackage{graphicx}
\usepackage{hyperref}  
\usepackage{xspace}
\usepackage{caption}
\usepackage{placeins}
\usepackage{xcolor}
\usepackage{subcaption}

\definecolor{afcolor}{HTML}{b3443c}

\begin{document}
\title{High-Redshift Galactic Outflows: Orientation Effects, Kinematics, and Metallicity in TNG50 and SERRA}

\author{Ivan Kostyuk\inst{1}, Stefano Carniani\inst{1}, Mahsa Kohandel\inst{1}, Andrea Pallottini\inst{2}}

\institute{Scuola Normale Superiore, Piazza dei Cavalieri 7,  50126 Pisa, Italy \\ \and
Dipartimento di Fisica ``Enrico Fermi'', Universit\'{a} di Pisa, Largo Bruno Pontecorvo 3, Pisa I-56127, Italy}

\abstract
{Galactic outflows driven by central black holes and supernovae play a crucial role in the formation and evolution of galaxies. Recently, JWST/NIRSpec observations have provided the first detections of warm ionised outflows in low-mass ($M_\star \sim 10^7 \mathrm{M}_\odot)$ galaxies at high redshifts ($z>3$), revealing an occurrence rate of 25-40\% depending on the intensity of the emission lines. This fraction is lower than predicted by theoretical models and simulations, which suggest that fast outflowing gas should be a common feature of all star-forming galaxies in the early Universe.}
{
In order to better understand the discrepancies between simulations and observations, we identify and characterize outflows in high-redshift galaxies using the TNG50 cosmological and SERRA zoom-in simulations. Our study examines how outflow detectability depends on the line of sight, explores the properties of the fast gas, and investigates its relationship with key galactic properties.}
{We analyse approximately $6\times10^{4}$ galaxies from TNG50 and $3\times10^{3}$ galaxies from SERRA over the redshift ranges $z=3$--$5$ and $z=4$--$5$, respectively, spanning stellar masses of $M_\star = 10^{7.5}$--$10^{11}\,\mathrm{M}_\odot$. Outflows in the immediate vicinity of each galaxy are identified using a Gaussian mixture model algorithm that uses the gas velocity, star-formation-rate, and location as input parameters. We subsequently compare the simulated outflows to those observed in the JWST/JADES NIRSpec survey.}
{Outflow masses in both TNG50 and SERRA broadly reproduce the JWST/JADES measurements within roughly 0.5 dex, though simulations tend to predict slightly higher values, suggesting that optical emission lines capture only a fraction of the multiphase outflow. However, simulated outflow velocities are typically an order of magnitude lower than those inferred from observations. TNG50 indicates a clear orientation dependence as outflows in face-on galaxies are approximately 15 percent more likely to be detected than in edge-on systems, with this difference increasing to nearly 40 percent for more massive, disc-shaped galaxies.}
{}
{}


\titlerunning{Characteristics of galactic outflows}
\authorrunning{Kostyuk et al.}
\maketitle

\section{Introduction}
Galactic outflows, primarily driven by supernova explosions and radiation feedback from active galactic nuclei (AGN), are expected to play a crucial role in regulating star formation and shaping the long-term evolution of galaxies (e.g. \citealp{debuhr12, li17, nelson19b, mitchell20, pandya21}). The general consensus is that the primary impact of outflows is the removal of gas from the galactic interior, leading to depletion of the star-forming reservoir. In massive galaxies, only the fastest outflows can escape the deep gravitational potential, while most of the gas eventually falls back and recycles into subsequent star formation. In contrast, lower-mass galaxies, with shallower potential wells, can lose a significant fraction of their gas, often resulting in early quenching of star formation (e.g. \citealt{tremoti04, christensen16}). Yet, despite significant progress, the role of outflows in shaping their host galaxies remains highly uncertain and debated.

Spectroscopic observations typically identify galactic outflows through broad components in the emission lines of a galaxy’s spectrum. Such broad components indicate the presence of fast-moving gas with velocities exceeding the circular velocity of the disk, suggesting that it may be able to escape into the IGM.
Several works have mainly studied the profile on the rest-frame optical line of H$\alpha$ and [OIII] that maps the ionised gas with a temperature of $10^{4-5}$~K (e.g. \citealp{arribas14, forster14, harrison16, cicone16, concas17, rakshit18, perna17, leung19, forster19, davies19, kakkad20, reichardt22, concas22, llerena23, rodriguez24}). In recent years, ground-based spectroscopic studies have detected outflows in high-mass galaxies ($M_\star > 10^{10}\mathrm{M}_\odot$) with an incidence rate exceeding 30\% at redshifts $z = 0 - 3$ (e.g. \citealp{carniani15, cicone16, concas17, rakshit18, forster19}).
Low-mass ($M_\star < 10^{10}\mathrm{M}_\odot$) galaxies at high redshifts ($z > 3$) remained difficult to study as their optical emission lines are redshifted into the near-infrared.
Understanding the outflow properties of these galaxies is crucial for assessing quenching mechanisms in the early Universe. 

The advent of the James Webb Space Telescope (JWST; \citealp{gardner06, gardner23}) with its Near Infrared Spectrograph (NIRSpec; \citealp{ferruit22, jakobsen22}) has provided the sensitivity needed for kinematic analysis of the optical lines at very high redshift, up to $z=9$, and to low stellar masses $\sim 10^7\mathrm{M}_\odot$ where the emission of the broad component is more challenging to observe from ground-based facilities.  Several studies, including those based on data from the Cosmic Evolution Early Release Science (CEERS; \citealp{bagley23, finkelstein23}), Early Release Observation (ERO; \citealp{pontoppidan22}), and GLASS-JWST Early Release Science (GLASS-JWST-ERS; \citealp{treu22}) surveys, have identified broad emission-line components likely originating from galactic outflows in low mass galaxies ($M_\star = 10^7 - 10^9\mathrm{M}_\odot$) \citep{tang23, zhang24, xu24}. Recent analysis of low-mass galaxies in a similar mass range from the JADES survey \citep{carniani24} has reported an outflow incidence rate of approximately 40\% depending on the strength of the emission line. The authors suggest that the lack of a ubiquitous detection of a broad component in all galaxies, as expected from cosmological simulations, is mainly due to the sensitivity of the observations and the projection effect of a biconical outflow along the line of sight. Clarifying the intrinsic incidence and orientation dependence of outflows is therefore essential for constraining the morphology of the observed outflowing gas.

The goals of this study are to investigate the physical factors that characterize galactic outflows at high redshift and to describe the mechanisms driving the observed incidence rate of outflows reported by \cite{carniani24} in JADES galaxies, thereby improving our understanding of their origin and properties. To this end, we analyse simulated galaxies from TNG50 and SERRA to explore how outflow detection depends on viewing angle, gas-phase metallicity, and stellar feedback strength, and to assess whether the relatively low occurrence rate found by JWST can be explained by orientation effects, physical conditions of the outflowing gas, or limitations in observational sensitivity.
We develop a novel approach based on a Gaussian mixture model to identify outflows in the immediate vicinity of galaxies, effectively distinguishing them from the galactic gas. This allows us to identify those outflows that would be directly observed in telescopes as opposed to complementary works e.g. \citep{nelson19b}, which primarily focus on outflows at $>10$kpc distance from the galactic center.
By utilizing two simulations built on fundamentally different frameworks, we aim to identify inconsistencies between the models themselves and to disentangle discrepancies with observations that arise from numerical or physical differences. This comparative approach provides deeper insights into the reliability of simulated outflow properties and their connection to real-world observations.

This paper is structured as follows: In Sect.~\ref{sec:Methodology}, we introduce the two simulations used in our study and describe the algorithm developed to identify galactic outflows. In Sect.~\ref{sec:Results}, we analyze the properties of these outflows and compare them to the observational data from the JADES survey. Finally, in Sect.~\ref{sec:Conclusions}, we summarize our findings and present our conclusions.

\section{Methodology \label{sec:Methodology}}

This section is divided into four parts. In Sec.\ref{subsec:tng50}, we describe the TNG50 cosmological simulation, and in Sec.\ref{subsec:serra}, the SERRA zoom-in simulation, both of which provide the data for our analysis. In Sec.~\ref{subsec:sample_selection} we describe the selection of our galaxy sample from each simulation. Finally, in Sec.~\ref{subsec:identfication}, we outline the methodology developed to separate outflowing gas from material associated with the galactic disc or inflows feeding the galaxy.

\subsection{The TNG50 simulation \label{subsec:tng50}}

The TNG50 simulation \citep{pillepich18b, springel18, naiman18, nelson18a, marinacci18} is a cosmological hydrodynamic simulation covering a comoving volume of $(51.7\mathrm{cMpc})^3$. Together with the TNG100 and TNG300 simulations, it is part of the IllustrisTNG project, which aims to provide detailed insights into galaxy formation and the processes driving it. TNG50 offers the highest resolution within this suite, achieving a baryonic mass resolution of $m_b = 8.5 \times 10^4 \mathrm{M}_\odot$ and a dark matter mass resolution of $m_\mathrm{dm} = 4.5 \times 10^5 \mathrm{M}_\odot$. The spatial resolution spans a broad dynamic range because the particle masses are nearly fixed: it reaches tens of parsecs in overdense regions such as galaxies, but is substantially lower in the diffuse intergalactic medium.

Gas dynamics within TNG50 is simulated using the AREPO code \citep{springel10}, which solves the equations of ideal magnetohydrodynamics (MHD) on a Voronoi mesh, allowing for accurate tracking of the gas evolution. The galaxy formation physics models follow the prescriptions developed by \cite{weinberger17} and \cite{pillepich18a}. Radiation feedback from stellar populations is not explicitly simulated. To prevent artificial gravitational collapse of dense, star-forming gas in the absence of such feedback, the subgrid model of \citet{springel03} is employed. This model maintains hydrodynamic stability by enforcing pressure equilibrium within the star-forming interstellar medium.

Star formation is modeled by stochastically converting gas cells that exceed a critical density threshold of $n_{\mathrm{H}} = 0.1\,\mathrm{cm}^{-3}$ into stellar particles according to the Kennicutt–Schmidt relation. Each stellar particle represents a stellar population that initially follows a Chabrier initial mass function (IMF) \citep{chabrier03}. Because the two-phase structure of the interstellar medium (ISM) is not explicitly resolved, supernova feedback would not be able to propagate through the low-density phase and would instead be artificially suppressed by a smooth ISM. To circumvent this, feedback from stellar particles is implemented via hydrodynamically decoupled kinetic wind particles \citep{pillepich18a}. These particles later recouple to the gas phase once they have left the dense ISM and the gas density has dropped below the critical threshold of $n_\mathrm{recouple} = 0.005\mathrm{cm}^{-3}$. While radiation transport is not explicitly simulated, a uniform ultraviolet (UV) background based on \citet{fg09} is activated at $z=6$. The simulation tracks chemical enrichment of gas, including the elements C, N, O, Ne, Mg, Si, Fe, and Eu individually.

Black holes are seeded into the centers of galaxies exceeding a halo mass of approximately $10^{10.8} \mathrm{M}_\odot$, with initial masses of $5 \times 10^5 \mathrm{M}_\odot$. Their subsequent growth is modeled via gas accretion following Bondi formalism, limited by the Eddington accretion rate, or through mergers between black holes. Feedback from black holes includes kinetic, thermal, and indirect radiative components, following the formalism developed by \citet{weinberger17}.

The TNG50 simulation uses cosmological parameters derived from the Planck 2016 measurements \citep{planck16}: $\Omega_m = 0.3089$, $\Omega_b = 0.0486$, $\Omega_\Lambda = 0.6911$, $h = 0.6774$, $\sigma_8 = 0.8159$, and $n_s = 0.9667$. Initial conditions are generated at a redshift of $z=127$ using the N-GenIC code \citep{springel05}, and the simulation is evolved down to $z=0$.

In our analysis, we retrieve galaxies from TNG50 catalogs, which have been identified using the SubFind algorithm \citep{springel01}, which employs the friends-of-friends method \citep{davis85} to detect gravitationally bound structures. Groups exceeding a particle-count threshold of 20 are classified as subhalos.

\subsection{The SERRA simulation \label{subsec:serra}}

The SERRA simulation \citep{pallottini22} comprises a suite of cosmological zoom-in simulations employing a significantly different numerical methodology compared to TNG50. SERRA utilizes the RAMSES adaptive mesh refinement (AMR) hydrodynamic code \citep{teyssier02} with an on-the-fly radiative transfer module \citep{rosdahl13}.
Radiation transport employs a momentum-based method with M1 closure and tracks five separate photon energy bins, with gas-radiation interactions that are coupled via KROME \citep{grassi14,pallottini:2017_b}, which incorporates a non-equilibrium chemical network tracking the abundances of species such as H, $\mathrm{H}^{+}$, $\mathrm{H}^{-}$, He, $\mathrm{He}^{+}$, $\mathrm{He}^{++}$, $\mathrm{H}_2$, and $\mathrm{H}_2^{+}$, and that is coupled with radiative transfer \citep{pallottini:2019}.
Star formation is modeled on a Kennicutt–Schmidt-like relation based on molecular hydrogen density, with stellar particles created stochastically following a Kroupa IMF \citep{kroupa01}.
Stars act as a source of mechanical, chemical, and radiative feedback according to PADOVA stellar tracks \citep{bertelli:1994} that is directly coupled with the gas \citep[see][for details]{pallottini:2017} without adopting an explicit 2-phase subgrid assumption.

Adopting Planck cosmological parameters, the MUSIC framework \citep{hahn11} was used to generate various sets of zoom-in initial conditions, with mass resolutions of up to $m_\mathrm{b} = 1.2 \times 10^4 \mathrm{M}_\odot$ for baryonic matter and $m_\mathrm{dm} = 7.2 \times 10^4 \mathrm{M}_\odot$ for dark matter, with spatial resolution reaching $\simeq 30\mathrm{pc}$ at $z=6$ in the highest-resolution regions, while being substantially lower in th diffuse intergalactic medium.

\subsection{Sample selection \label{subsec:sample_selection}}

For our analysis of TNG50 galaxies, we select all systems with stellar masses $M_\star > 10^{7.5}\,\mathrm{M}_\odot$ within the redshift range $3 \leq z \leq 6$, yielding a sample of approximately 60,000 galaxies. At this mass scale, the simulation provides a sufficient numerical resolution of $\gtrsim 10^4$ gas particles per galaxy.

For the SERRA simulation, we focus on zoom-in regions targeting the most massive haloes ($M_{\mathrm{h}} \geq 10^{11}\,\mathrm{M}_\odot$) and their satellite populations, with volumes extending to roughly six times each halo's virial radius. In this study, we include galaxies evolved down to $z = 4$ \citep[see][for the sample]{kohandel:2024}. From these zoom-ins, we extract all galaxies with halo masses $M_{\mathrm{h}} \geq 10^{9}\,\mathrm{M}_\odot$, corresponding to typical stellar masses of $M_\star \gtrsim 10^{7.5}\,\mathrm{M}_\odot$ within the redshift interval $4 \leq z \leq 6$. This selection yields a sample of about 3,000 SERRA galaxies.

\begin{figure}
    \centering
    \includegraphics[width=\linewidth]{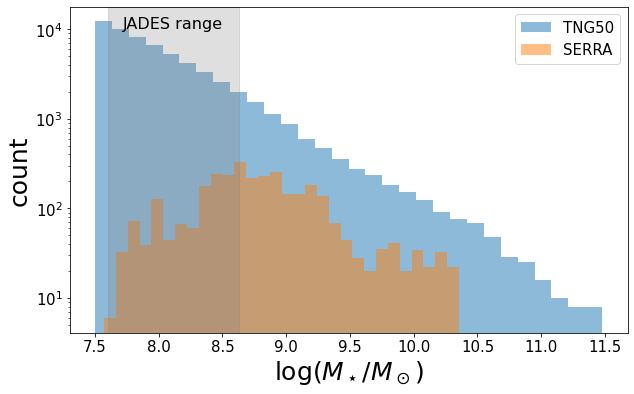}
    \caption{Number of galaxies as a function of stellar mass analysed in this work. The blue and orange histograms report the distribution for TNG50 and SERRA, respectively. The shaded region indicated the mass range of galaxies observed in JADES.}
    \label{fig:sample_counts}
\end{figure}

In both simulations, the selected galaxies do not represent independent populations but, to a certain extent, trace the same systems observed at different cosmic epochs. Because the SERRA zoom-ins are centered on massive halos, their galaxy populations are dominated by satellites residing in highly overdense environments. Fig.~\ref{fig:sample_counts} shows the number of galaxies in each sample as a function of stellar mass. Note that TNG50 provides a representative sampling of the cosmic web, whereas the SERRA sample is intrinsically biased toward dense regions surrounding massive central halos.

\subsection{Identification of outflows \label{subsec:identfication}}

The main difference between the outflow selection in \citet{nelson18a} and our method is that the former estimates outflows using the gas flux at a fixed radius. In contrast, our approach is designed to mimic the observational apertures within which the [O \textsc{iii}] and H$\alpha$ lines are measured. Therefore, we focus on outflows located close to, and partially intertwined with, the galaxy itself, whereas \citet{nelson18a} specifically select outflows at distances of 10–20 kpc from the galactic centre. Hence, the main challenge of our approach is to distinguishing gas particles associated with galactic material from those constituting outflows. To address this, we developed a separation algorithm employing a multivariate Gaussian mixture model combined with specific physical criteria.

Initially, for each TNG50 galaxy, we define a cut-out radius of 5 times the star-formation radius\footnote{The star formation radius is defined at the radius at which half of the galaxy's SFR is reached. This scale is essentially arbitrary, it only needs to ensure that the galaxy with most of its outflows is included, while not too large to avoid including additional galaxies in the neighborhood.} $r_\mathrm{SFR}$ for $M_\star<10^9\mathrm{M}_\odot$ galaxies and 10 times the star-formation radius for more massive galaxies, as these tend to be more compact. For galaxies from the SERRA simulation, the cut-out scale was selected to be twice the size of the field of view, with a radius of $0.6''$\footnote{Here the choice was made differently as SERRA galaxies are less diffuse, such that the $r_\mathrm{SFR}$ sets a scale that is too small.}. The different choices are made to ensure that most of the outflows are contained within the analysed volume while preferably avoiding neighbouring galaxies.

As a first step in our outflow selection process, we select gas particles exhibiting outward radial velocities relative to the galactic centre, defined by the condition $v_\mathrm{r} = \vec{v}\cdot \vec{r}_0 > 0$, where $\vec{r}_0$ is the normalized position vector of the particle relative to the galaxy centre, and $\vec{v}$ is its velocity vector relative to the galactic centre. This essentially sets a first prior for our selection algorithm.

Subsequently, we distribute the selected particles into a five-dimensional space spanned by their spatial coordinates ($\vec{x_i}$), outward radial velocities ($v_r$), and star formation rate (SFR).
The choice for this space is based on the assumption that outflowing gas typically exhibits higher outward velocities, has a different distribution (i.e., larger scales) and morphology (i.e, bi-conical) as compared to the gas that makes up the galaxy, and displays lower star-formation activity compared to galactic gas.
Based on these assumptions, we classify gas particles using a Gaussian mixture model \citep{dempster71}. We model the distribution of particles in the aforementioned 5-dimensional space with a multimodal Gaussian profile defined by:
\begin{equation} \label{eq: MixtureModel}
p(\vec{r}, v_\mathrm{r}, \mathrm{SFR}) = \sum_{i=1}^{n_\mathrm{modes}} \phi_i \mathcal{N}(\vec{\mu}_i, \Sigma_i),
\end{equation}
where $\phi_i$ represents the weight of the $i$-th Gaussian mode out of a total of $n_\mathrm{modes}$ modes, $\mathcal{N}$ characterized by mean\footnote{In our case $\vec{\mu_i} = (\vec{x_i}, v_r, \mathrm{SFR})$.} $\vec{\mu}_i$ and covariance matrix $\Sigma_i$. First, we determine the maximum likelihood configuration of the aforementioned probability distribution to best fit the distribution of gas particles. Subsequently, each gas particle is assigned to the mode to which it has the highest likelihood of belonging. 

To determine the number of modes, we inspected a sample of galaxies and attempted to separate the outflows by using equation \ref{eq: MixtureModel} with different values of $n_\mathrm{modes}$. We performed visual inspections of galaxy samples to guide the classification. We found that $n_\mathrm{modes} = 3$ to be the most effective choice, as a higher number of modes often resulted in the galaxy being separated into more than one mode, while having only two modes frequently led to the inclusion of outflows close to the galaxy into the galactic mode. 
We note that a model with three Gaussian modes usually associates one mode with all particles closer to the galaxy centre, and we assign this mode to the gas in the galaxy that has a random motion. The other two modes are attributed to the outflow component.

Finally, as an additional selection criterion, we consider gas particles that, while not necessarily having positive radial velocities and thus do not satisfy our initial prior, exhibit nonetheless substantial velocities perpendicular to the galactic plane\footnote{By galactic plane, we refer to the plane perpendicular to the angular momentum vector of the galaxy. Thus, this definition also includes low-mass diffuse galaxies which do not necessarily have a disc structure.}. This additional criterion is motivated by the possibility of gas being launched tangentially, resulting in minimal radial velocity. We thus introduce a second outflow identification criterion based on the velocity component perpendicular to the galactic plane:
\begin{equation}
\text{Outflow} =
\begin{cases}
v_z > 2\sigma & \text{if } z \geq 0 \\
v_z < -2\sigma & \text{if } z < 0,
\end{cases}
\end{equation}
where $\sigma$ denotes the one dimensional velocity dispersion\footnote{While for TNG50 $\sigma$ is provided in the data release, for SERRA $\sigma$ is computed by taking the square root of the variance of the velocity of the gas located within 2.5 kpc from the galactic center.}. This criterion ensures that the outflows selected this way are statistically unlikely to belong to the galaxy.

\begin{figure*}
    \centering

    \begin{subfigure}{0.48\linewidth}
        \includegraphics[width=0.48\linewidth]{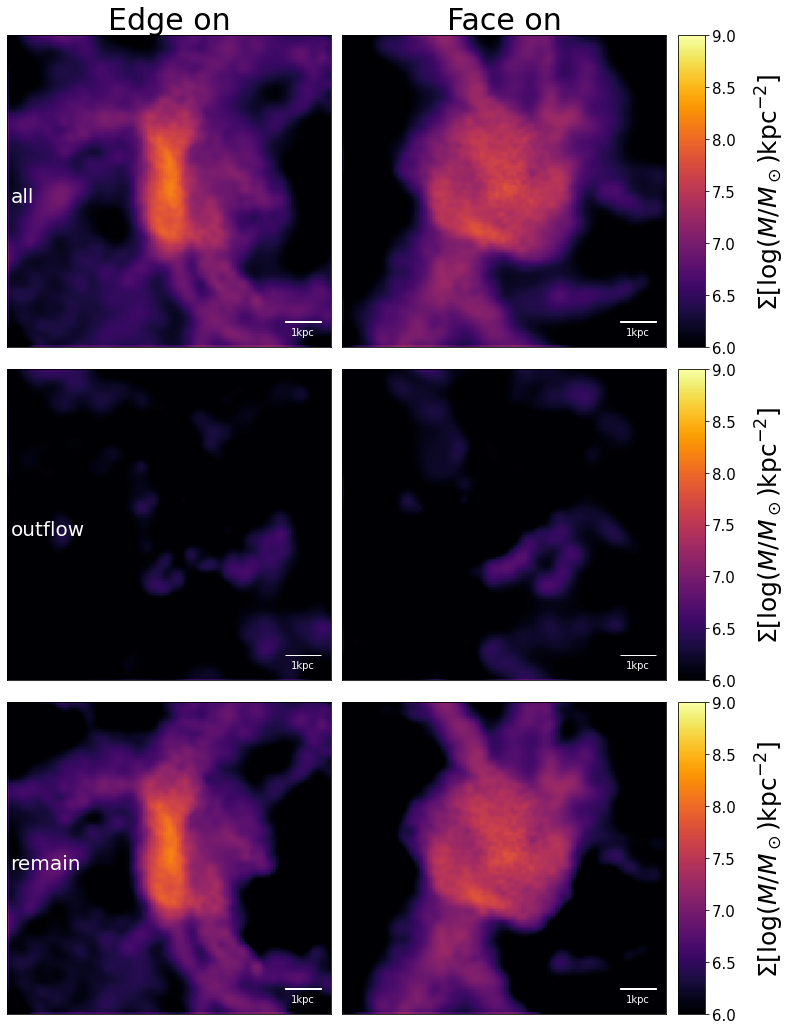}
        \includegraphics[width=0.48\linewidth]{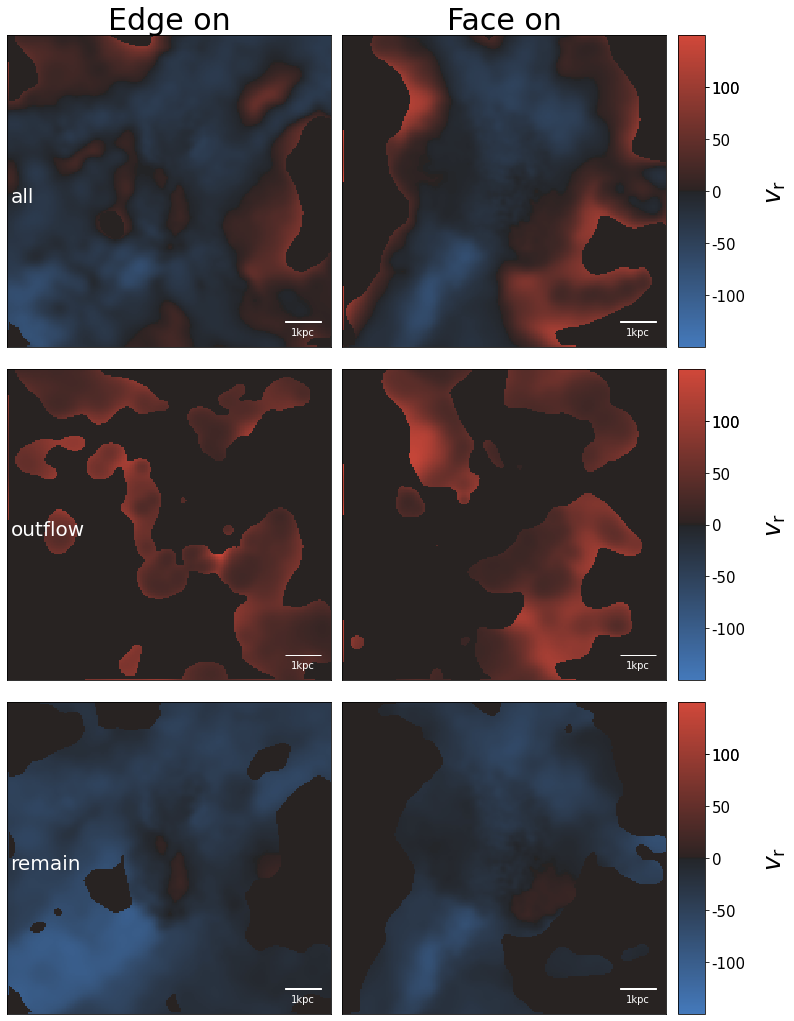}
        \caption{TNG50 $M=10^8\mathrm{M}_\odot$ at $z=3.3$}
    \end{subfigure}
    \begin{subfigure}{0.48\linewidth}
        \includegraphics[width=0.48\linewidth]{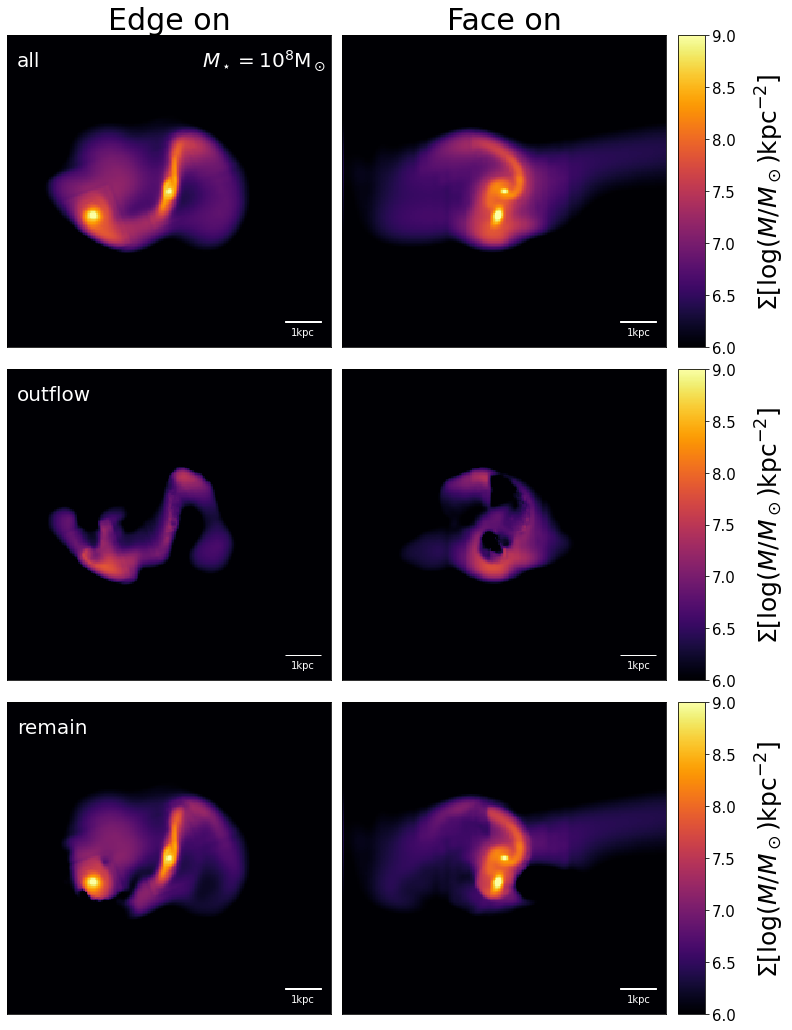}
        \includegraphics[width=0.48\linewidth]{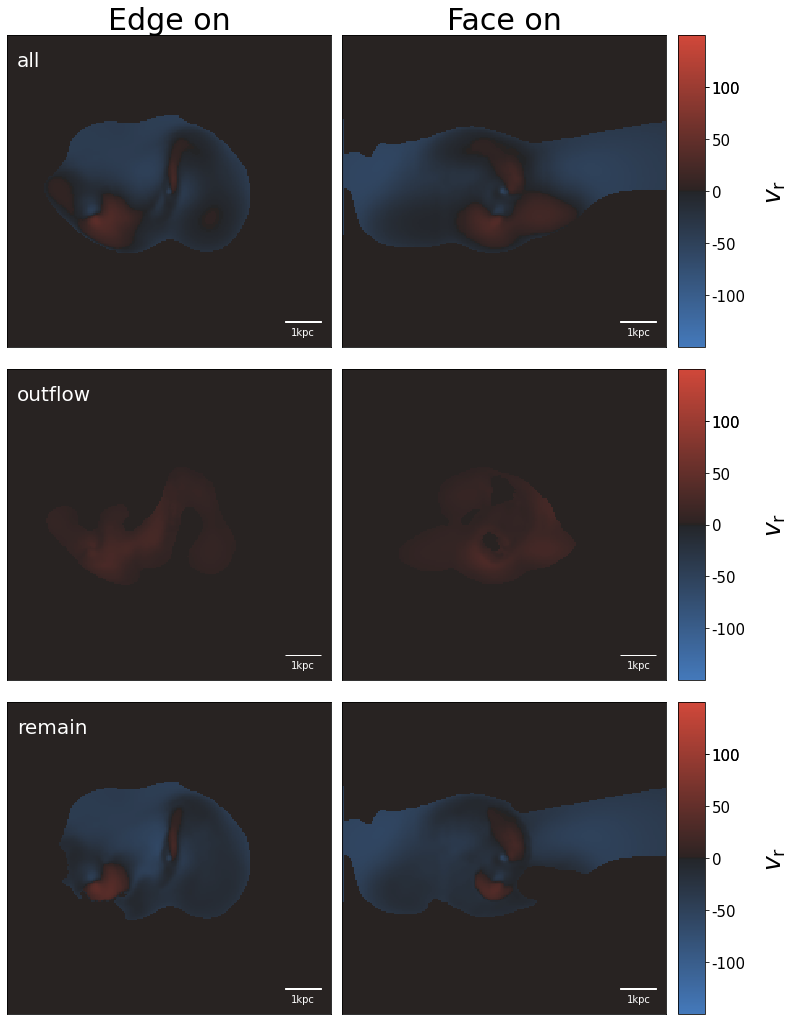}
        \caption{TNG50 $M=5\times 10^{10}\mathrm{M}_\odot$ at $z=3.3$}
    \end{subfigure}

    \begin{subfigure}{0.48\linewidth}
        \includegraphics[width=0.48\linewidth]{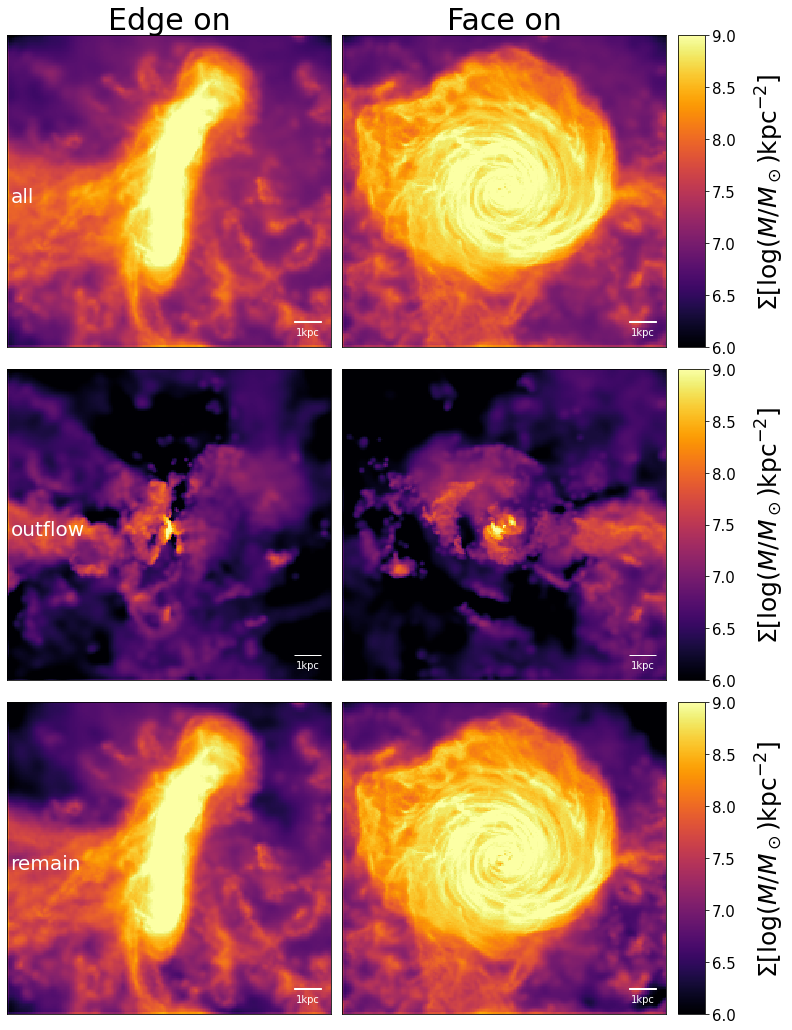}
        \includegraphics[width=0.48\linewidth]{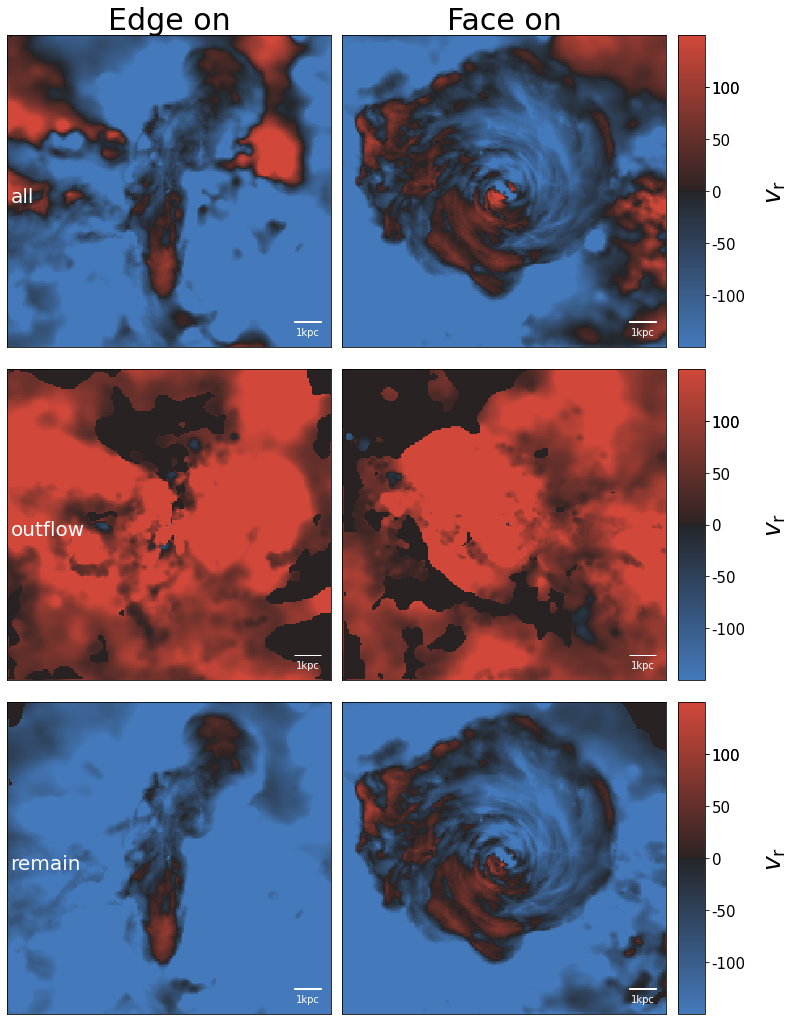}
        \caption{SERRA $M=10^8\mathrm{M}_\odot$ at $z=5$}
    \end{subfigure}
    \begin{subfigure}{0.48\linewidth}
        \includegraphics[width=0.48\linewidth]{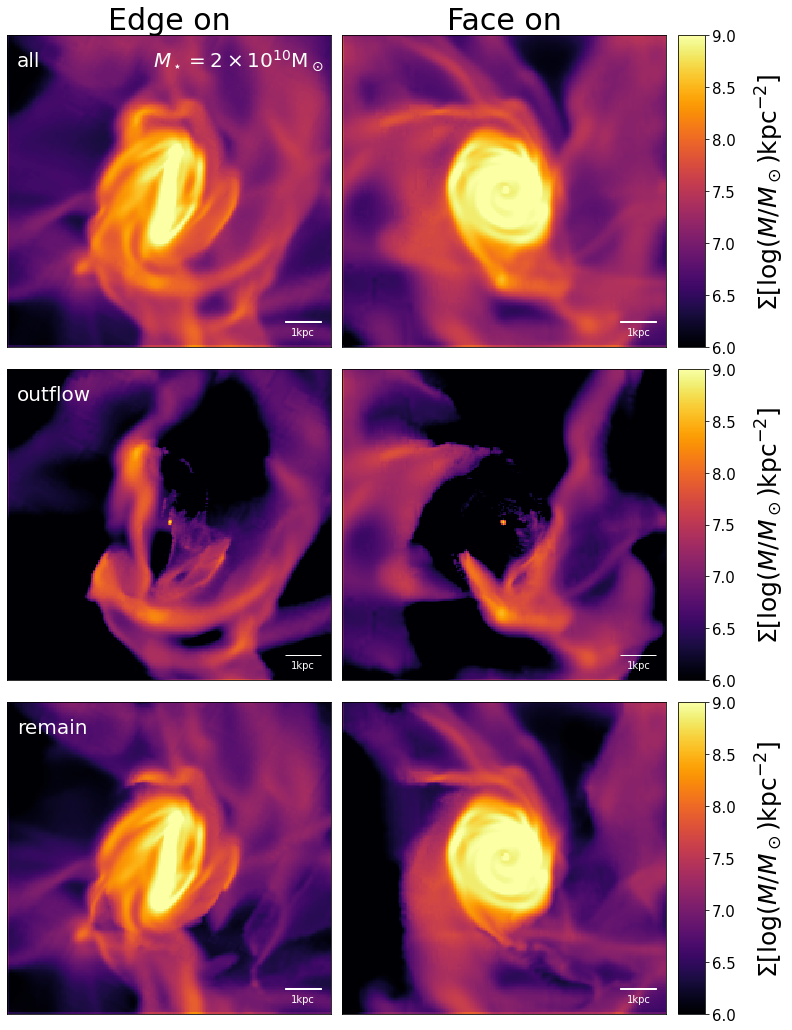}
        \includegraphics[width=0.48\linewidth]{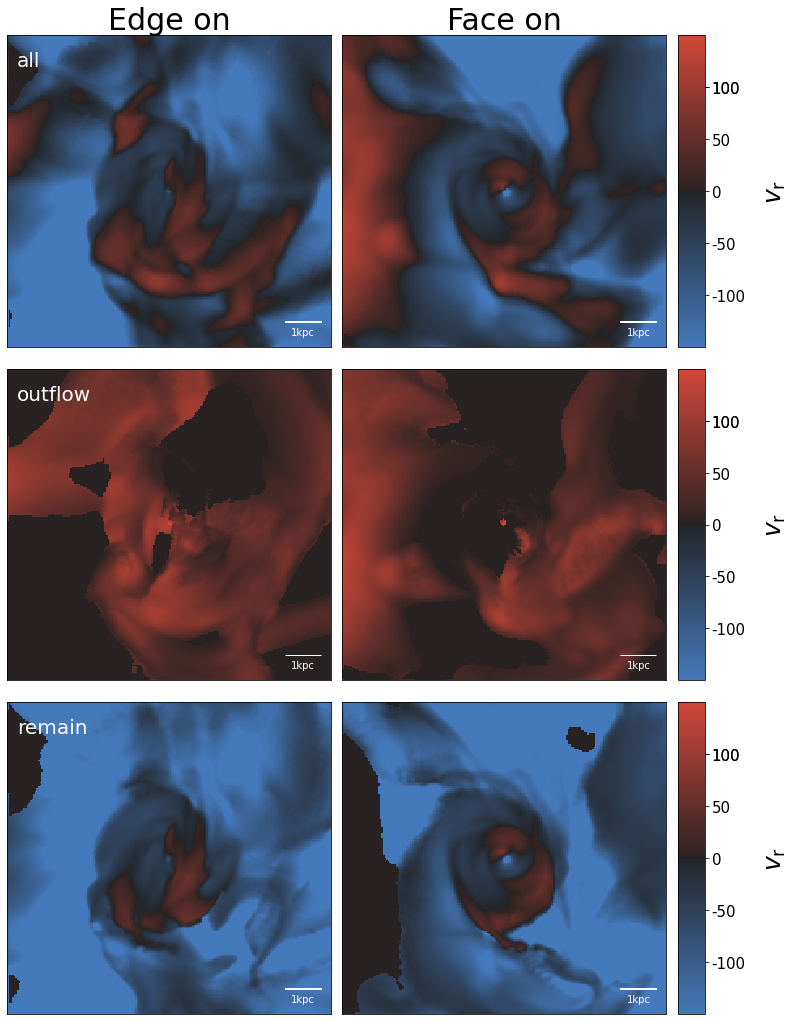}
        \caption{SERRA $M=2\times 10^{10}\mathrm{M}_\odot$ at $z=4.2$}
    \end{subfigure}

    \caption{Two sample galaxies from the TNG50 (left) and SERRA (right) simulations respectively, shown in edge-on and face-on view together with their corresponding radial velocity maps. From top to bottom, each column shows the full projected gas of the galaxy, the fraction of gas identified as outflows, and the fraction of gas associated with the galaxy itself.}
    \label{fig:example_galaxies}
\end{figure*}

Fig.~\ref{fig:example_galaxies} illustrates our outflow selection process for four representative galaxies. Two from the TNG50 simulation and two form the SERRA. For each simulation we depict a low mass galaxy with $M_\star = 10^8 \mathrm{M}_\odot$ and a high mass galaxy with $M_\star = 2-5\times 10^{10} \mathrm{M}_\odot$. 

For TNG50, we see that the lower mass galaxy is significantly more diffuse and without a clearly discernible disc structure. The higher mass galaxy, on the other hand, has a clear disc structure, and we see a clear outflow jet originating from the galactic centre.



In SERRA, the lower-mass galaxy on the left is undergoing a merger. In fact, we have found that about $60\%$ of our sample of SERRA galaxies is in the process of merging. As shown, the extraction algorithm attributes part of the companion galaxy to the outflows. This is reasonable, as the tidal arm and portions of the galactic disc exhibit significant outward velocities relative to the central galaxy. Such kinematic features, as discussed in \citep{kohandel19}, would likely appear as a broad component in observational data, effectively mimicking what telescopes would detect.

By analysing TNG50 and SERRA simulations, we note that the latter tend to form disc-like structures at lower masses ($M_\star<10^8 \mathrm{M}_\odot$) and earlier times with respect to TNG50 \citep[compare with][]{kohandel:2024}. As a result, the gas in SERRA galaxies is more clumpy and less diffuse than the gas in the TNG50 counterparts of similar mass, such as the example on the left in Fig.~\ref {fig:example_galaxies} \citep[see also][for similar comparison]{fujimoto:2025}. The galactic morphology of higher-mass ($M_\star>10^{10} \mathrm{M}_\odot$) galaxies in SERRA seems more comparable to those in the TNG50 simulation. 


\section{Results \label{sec:Results}}

\subsection{General outflow properties}

\begin{figure}
    \centering
    \includegraphics[width=\linewidth]{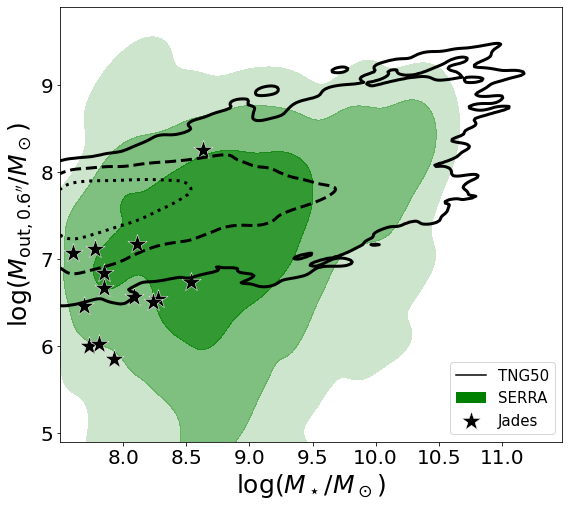}
    \caption{Mass of outflowing gas as a function of galactic stellar mass, measured within a radius of $0.6"$. The green contours represent the 1, 2, and 3$\sigma$ distribution of SERRA galaxies, while the black contours represent the same for TNG50 galaxies. Black stars represent galaxy observations from JADES.}
    \label{fig:Mout}
\end{figure}

Once outflows are identified in simulations, we estimate their physical properties and compare them with the observations. To ensure a fair comparison with observational constraints, we compute outflow masses within an area of $0.6''$, corresponding to a physical radius of $r=3.5-4.7$kpc, from the galactic center to mimic the spatial coverage of the JWST/NIRSpec MOS observations\citep{jakobsen2022}. Fig.~\ref{fig:Mout} shows the distribution of galaxies in our sample as a function of stellar mass and associated outflow mass. We note a mild correlation between outflow and stellar masses in both sets of simulations. The outflow mass estimates of both TNG50 and SERRA are broadly consistent with JADES observations by \cite{carniani24}. However, TNG50 outflows are on average $\sim0.5$ dex higher than those estimated from observations. 

This offset might indicate that JADES may capture only a fraction ($\sim35\%$) of the total outflowing material. The discrepancy likely reflects the multiphase nature of outflows: our model accounts for gas across all phases and temperatures, whereas H$\alpha$ or [O III] emission maps isolate only the warm ionized component at $10^{4-5}~{\rm K}$. On the other hand, the observations seem to be consistent with the predictions from SERRA, indicating that the optical lines probe a large fraction of gas in the outflows expected from SERRA.
A subset of SERRA galaxies exhibits outflow masses lower than $10^{6.5}~{\rm \mathrm{M}_\odot}$ that are not seen in the TNG50 simulation. This difference is likely driven by the different feedback implementations: while in TNG50 the star formation is mostly suppressed via outflows, in SERRA it is more gently regulated via preventative feedback, i.e. molecular hydrogen photodissociation and turbulence injection (see \citealt{pallotini25} for a more complete discussion), thus large-scale winds are expected to be less prominent.

\begin{figure*}
    \centering
    \begin{subfigure}{0.33\linewidth}
        \centering
        \includegraphics[width=\linewidth]{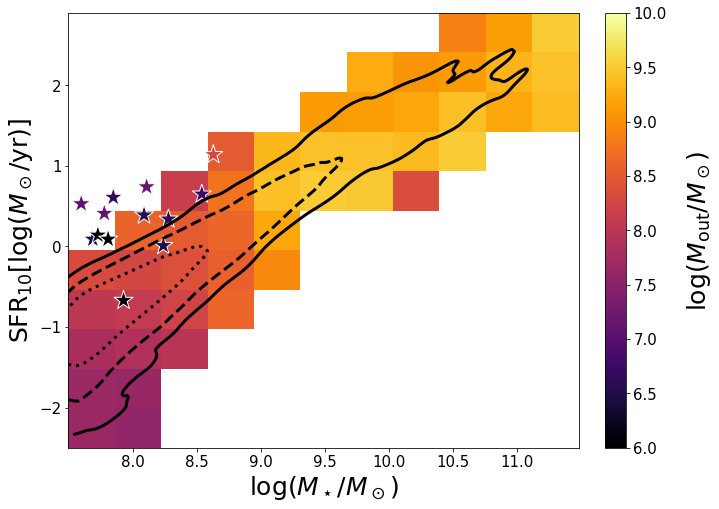}
        \subcaption{TNG50 total outflows}
    \end{subfigure}
    \begin{subfigure}{0.33\linewidth}
        \centering
        \includegraphics[width=\linewidth]{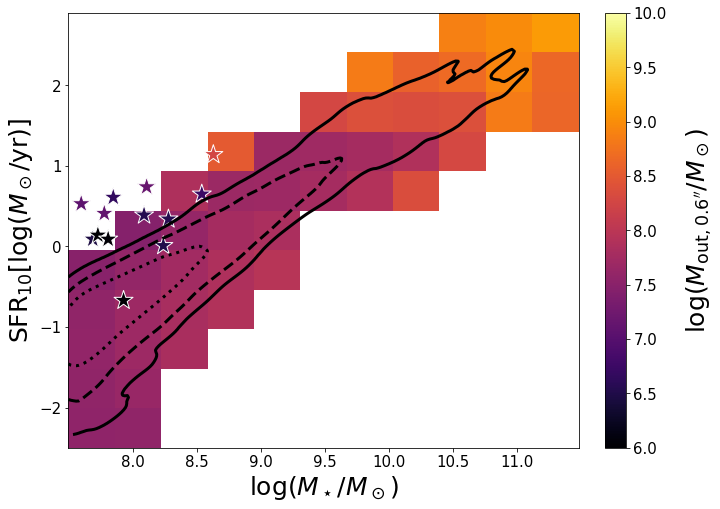}
        \subcaption{TNG50 aperture outflows}
    \end{subfigure}
    \begin{subfigure}{0.33\linewidth}
        \centering
        \includegraphics[width=\linewidth]{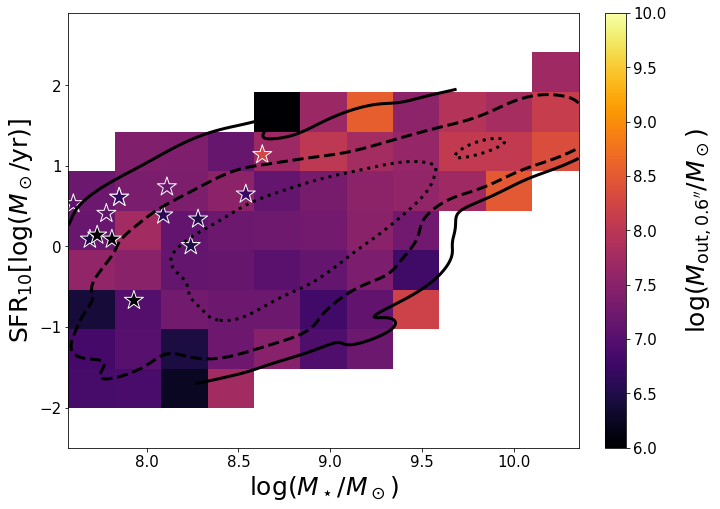}
        \subcaption{SERRA aperture outflows}
    \end{subfigure}

    \caption{Star formation rate as a function of stellar mass. The color bar indicates the average mass of outflowing gas in each bin for the total outflowing gas of TNG50 (left) galaxies, the outflowing gas within the aperture of TNG50 galaxies (centre), and the corresponding values for SERRA galaxies (right). Overplotted stars represent galaxies observed by JADES, with their colours representing the outflow mass following the same scale as the respective panels.}
    \label{fig:SFR}
\end{figure*}

Fig.~\ref{fig:SFR} depicts the SFR-$M_\star$ relations for TNG50, SERRA together with the JADES observations. Symbols and distribution are colour-coded by the measured outflow mass. The left panel shows the total outflow mass of TNG50 galaxies, the central panel restricts this to outflows within the aperture, while the right panel presents the aperture-limited outflows of SERRA galaxies. For the latter an extended field of view is not informative, as in almost all cases this would include additional galaxies contaminating the data. Intuitively, the outflow mass is expected to scale with the 10 Myr averaged SFR at fixed stellar mass. This is in particular true for lower-mass galaxies as in this case outflows are predominantly powered by supernovae from massive stars, which typically occur a few Myr after their formation. While this should holds in a quasi-steady state situation, if burstiness becomes important on a timescale different than 10 Myr \citep{pallottini:2023, sun:2025}, a decoupling can be expected between SFR and subsequent outflows.

The observed galaxies have an approximately 1.5 orders of magnitude higher SFR with respect to the average SFR of the mass-matched galaxies from TNG50 and are thus located outside the 3$\sigma$ contour. Because TNG50 is calibrated to reproduce the observed stellar masses at redshift zero, the total star-formation rate integrated over cosmic time is correct by construction. A mismatch at higher redshift therefore points to two possibilities. Either the evolution of the star-formation history differs from the model, meaning that star formation occurs at different epochs than TNG predicts, or, more plausibly, burstiness is higher than what the model predicts. Indeed, increased burstiness would naturally broaden the range of SFR values within each stellar-mass bin. Additionally, this result could suggest that observations are biased towards detecting outflows only in starburst galaxies driving larger outflows.

In contrast to the TNG50 results, the SFRs seen in the SERRA sample align well with the JADES observations. The observations are still located between 1-3$\sigma$ from the average value. However, unlike TNG50, SERRA is not calibrated to reproduce the stellar mass function at redshift zero. Moreover, the plot seems to slightly indicate that galaxies with higher SFRs tend to exhibit greater outflow masses. However, no strong correlation is visible, likely due to a substantial portion of the outflows being associated with tidal features from merger events rather than being driven by stellar feedback.

When investigating the total outflows in TNG50 (left plot in Fig.~\ref{fig:SFR}), galaxies with low stellar masses ($M_\star \lesssim 10^9 \mathrm{M}_\odot$), the total mass of outflowing gas correlates with the SFR at fixed stellar mass, whereas this correlation weakens or disappears at higher stellar masses. As discussed in Appendix~\ref{app:wind}, this is mainly caused by the fact that a significant fraction of the outflows in higher-mass galaxies is contained in the hydrodynamically decoupled wind particles. Once these particles are included (see Fig.~\ref{fig:wind_sfr}) the correlation is also clearly seen in more massive galaxies.

When limiting the outflowing mass to only the aperture, the correlation for low mass galaxies with SFR vanishes while a slight correlation can be seen for galaxies with $M_\star>10^{9.5}\mathrm{M}_\odot$. This likely arises due to the fact that low mass galaxies are highly diffuse in TNG50 (see Fig.~\ref{fig:example_galaxies}) such that even at larger radii the gas density remains relatively high. Given that wind particles only hydrodynamically recouple to the surrounding medium at a threshold density of $0.005\mathrm{cm}^{-3}$ (see Sec.~\ref{subsec:tng50}), some of the outflows are missed at closer distances as they only recouple later on.

\begin{figure*}
    \centering
    \includegraphics[width=0.49\linewidth]{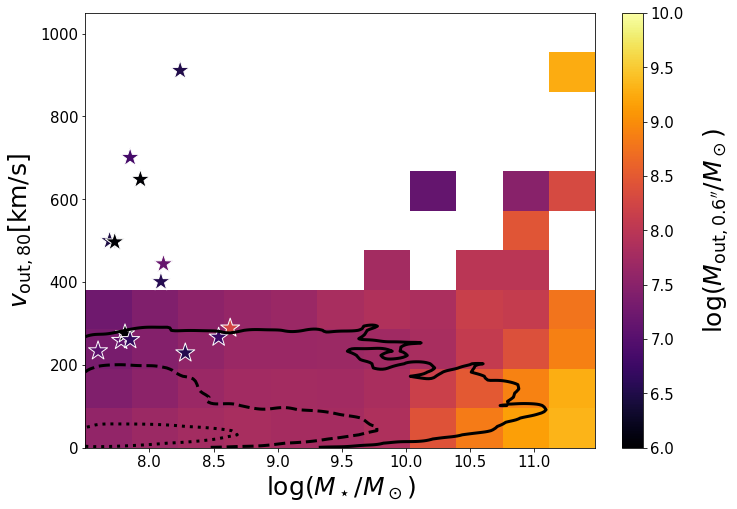}
    \includegraphics[width=0.49\linewidth]{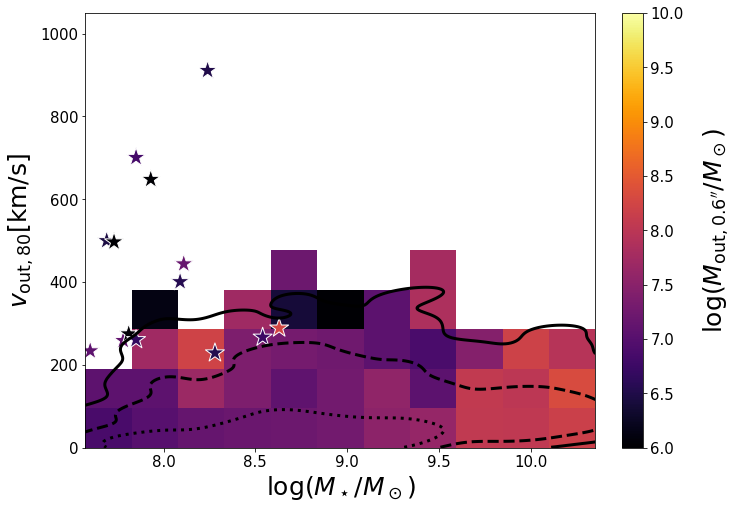}
    \caption{Velocity of the 80th percentile of outflowing gas as a function of stellar mass. Velocities are measured within a radius corresponding to an observed radius of $0.6'' $ for TNG50 galaxies (left) and SERRA galaxies (right). The colour bar indicates the average velocity value in each bin. Stars represent galaxies observed by JADES, with colours following the same scale as the histograms.}
    \label{fig:vout}
\end{figure*}

In observations, outflow velocities are typically inferred from the line width of the component associated with fast gas. For example, \cite{carniani24} measured the outflow velocity as $v_{\rm out} = |v_{\rm broad}-v_{\rm narrow}| + 2\sigma_{\rm broad}$ where $|v_{\rm broad}-v_{\rm narrow}|$ is the velocity shift between the peak of the broad and narrow emission line components, and $\sigma_{\rm broad}$ is the velocity dispersion of the broad component. The results in the work seem to suggest that this measurement should weakly depend on the outflow orientation with respect to the line of sight. In simulations, however, there is no unique definition for the outflow velocity, as each particle has its own value. By studying the distribution of radial velocities of outflowing particles, we find that the range varies significantly across targets. We therefore adopt an approach similar to that used in observational studies and define the outflow velocity as the 80th percentile of the radial velocity distribution, hereafter denoted as $v_{\mathrm{r},80}$.
%
In Fig.~\ref{fig:vout}, we investigate the outflow velocity $v_{\mathrm{r},80}$ as a function of galactic stellar mass. The colour scale indicates the average outflow mass within each bin, revealing an anti-correlation between outflow velocity and outflow mass for TNG50 galaxies, while no noticeable correlation can be seen for SERRA galaxies. Two potential explanations could account for this trend. First, physically, if a fixed amount of energy from supernovae is distributed among a larger mass of gas, the resulting average kinetic energy per gas particle will be lower. Alternatively, the observed trend might stem from a numerical artifact, whereby higher outflow masses result from the misclassification of galactic gas as outflowing material, which in turn tends to be slower.

In both cases, we find that most JADES galaxies exhibit far higher outflow velocities than those found in either TNG50 or SERRA. In TNG50, outflow velocities are determined by the specific implementation of wind particles (see Sec.~\ref{subsec:tng50}). In particular, their velocities are prescribed to scale with the host halo’s velocity dispersion, although this relationship may not hold in reality or may deviate from a strictly linear scaling.

\begin{figure}
    \centering
    \includegraphics[width=\linewidth]{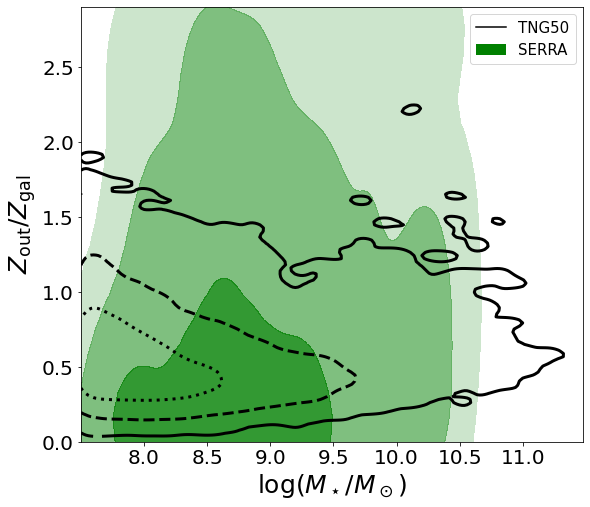}
    \caption{Ratio between the metallicity of the outflowing gas and the metallicity of the galactic gas as a function of galactic stellar mass, measured within a radius of $0.6"$. The green contours represent the 1, 2, and 3 $\sigma$ distribution
of SERRA galaxies, while the black contours represent the same for TNG50 galaxies.}
    \label{fig:Zratio}
\end{figure}

Finally, in Fig.~\ref{fig:Zratio}, we investigate the metallicities of the outflows by examining the ratio of the metallicity of the outflowing gas, $Z_\mathrm{out}$, to that of the galactic gas, $Z_\mathrm{gal}$. We observe that outflows typically exhibit significantly lower metallicities compared to their host galaxies. For TNG the average metallicity ratio is $\langle Z_\mathrm{out}/Z_\mathrm{gal} \rangle_\mathrm{TNG} = 0.52$ and for SERRA the ratio is even lower at $\langle Z_\mathrm{out}/Z_\mathrm{gal} \rangle_\mathrm{SERRA} = 0.16$.
Several factors might explain this difference. 

Firstly, outflows originate in regions of intense star formation, where the gas and stellar populations are younger and thus less enriched with metals. 
Secondly, supernova-driven outflows can sweep up gas that lies outside the star-forming regions or even gas still undergoing inflow, which has not yet mixed thoroughly with metal-enriched gas.
Thirdly, in the TNG50 simulation, the mass loading due to star formation is scaled down with the metallicity of the star-forming region based on the idea that high metallicity regions are more efficient at dissipating energy through radiation \citep{pillepich18a}.
Fourth, if the feature identified as an outflow actually corresponds to a tidal tail generated during a merger event, as is often the case in SERRA, the observed low metallicity may be explained by the presence of accreting gas that has not yet undergone star formation.

An important implication of this result is that the mass of outflows may be systematically underestimated when metal emission lines such as [O III] are used as tracers. Observationally, the [O III] luminosity is converted into an outflow mass by adopting an assumed oxygen abundance. This introduces a bias if the outflowing gas does not, in fact, share the same metallicity as the host galaxy.

\subsection{Orientation dependence}

In this section, we examine how galaxy orientation influences the detectability of galactic outflows.
From an observational point of view, outflows are typically identified by the presence of a broad component in the emission line profile, which indicates high-velocity gas along the line of sight. If the outflowing gas moves predominantly perpendicular to the line of sight the projected velocity is close to zero. As a result, the spectral signature lacks the broadened wings associated with fast-moving gas, making the outflow much harder to detect (see e.g. \citealt{phillips25}).
Galaxies that are converging toward a disc-like morphology are expected to exhibit outflows that are more easily detectable in face-on orientations than in edge-on views. This is due to several factors. First, the gas density perpendicular to the galactic disc is typically lower, providing less resistance to escaping material and allowing outflows to develop more prominently. Second, because there is no rotational component in the direction perpendicular to the disc, the gas velocity component in this direction is generally lower than within the plane, which makes it easier to identify outflowing components. Third, since star formation is concentrated in galactic disc, outflows are often launched preferentially along the minor axis, perpendicular to the plane, where they are more visible in face-on orientations.

\cite{carniani24} and \cite{xu24} suggest that the relatively low incidence rate of outflows—around 30\% in high-$z$ galaxies may be influenced by galaxy inclination.
To investigate this possibility, we projected the velocities of both ``outflow'' and ``remain" gas particles along various lines of sight relative to the plane of the galaxy, defined as the plane perpendicular to its angular momentum vector. The distribution of projected velocities is expected to approximate the spectral profile of the emission line produced by the same gas particles.
We stress, however, that the observed emission-line profile matches the velocity distribution only under the assumption that all gas particles emit with luminosity proportional to their mass\footnote{A more detailed analysis will be presented future work, where mock rest-frame optical lines will be generated. Some analysis of indicidual SERRA galaxies is presented in \cite{kohandel:2025} and \cite{phillips25}.}. We then calculated the width of the distribution using $W_{80}$, defined as the difference between the velocities at the 90th and 10th percentiles. For a Gaussian profile, $W_{80}$ is comparable to the FWHM.

\begin{figure*}
    \centering
    \includegraphics[width=0.48\linewidth]{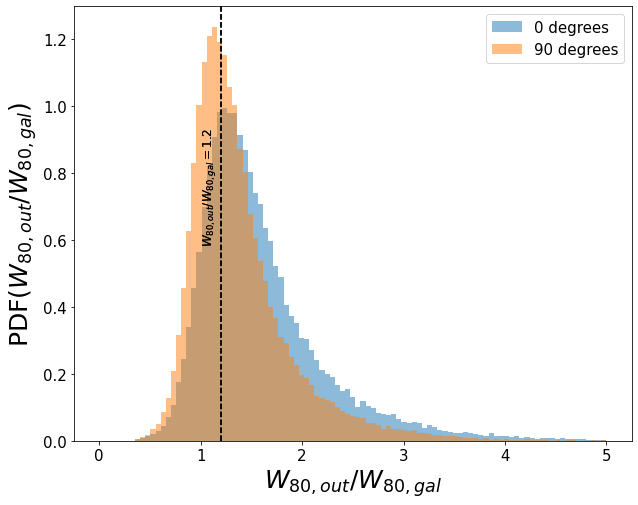}
    \includegraphics[width=0.48\linewidth]{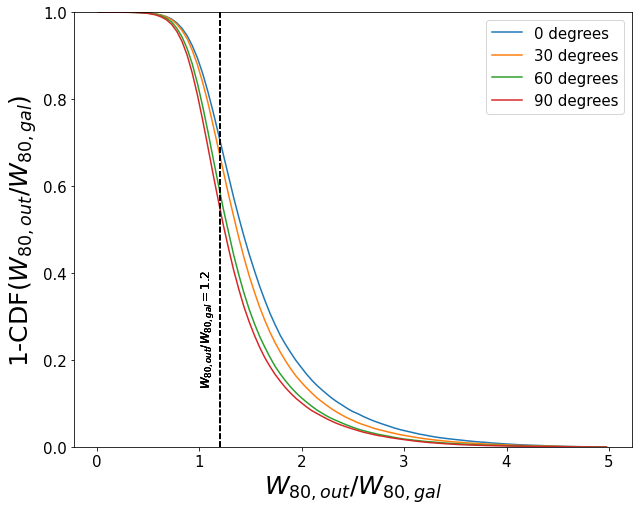}
    \includegraphics[width=0.48\linewidth]{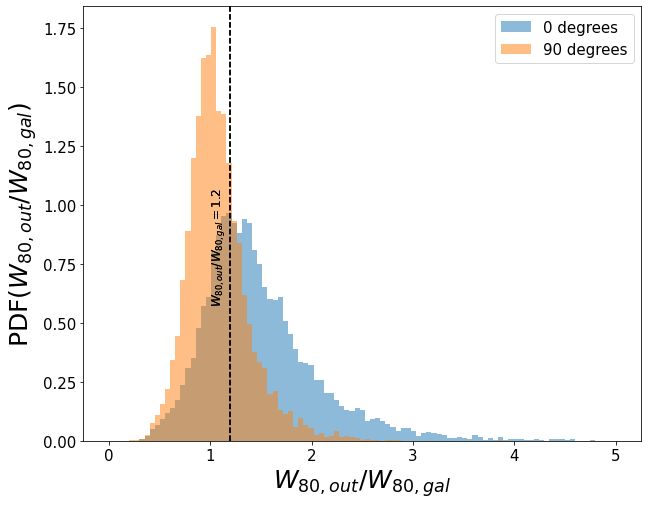}
    \includegraphics[width=0.48\linewidth]{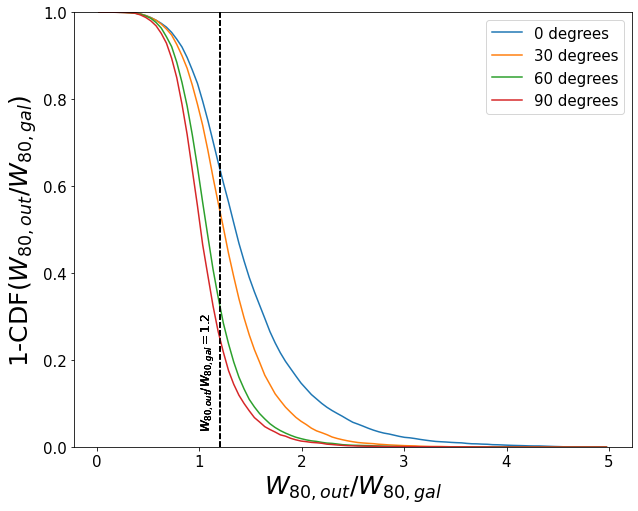}
    \caption{Top-left: Distribution of the ratios of velocity distribution widths between the outflows ($W_{80, \mathrm{out}}$) and the galaxy ($W_{80, \mathrm{gal}}$) in TNG50 galaxies for a face-on orientation (0$^\circ$) and an edge-on orientation (90$^\circ$). The ratio of 1.2, indicated by the vertical line, represents minimal ratio used by JADES to separate outflows from galactic gas. \\
Top-right: Complement cumulative distribution of the width ratios for various galactic orientation angles. \\ Bottom: Same as above but limited to galaxies with $M_\star>10^{8.5}\mathrm{M}_\odot$.}
\label{fig:orientation}
\end{figure*}

Initially, we analysed the results from the TNG50 simulation.
The left panels of Fig.~\ref{fig:orientation} show the distribution of ratios of the $W_{80, \mathrm{out}}$ of the ``outflow'' gas and $W_{80, \mathrm{gal}}$ of the ``remain'' gas particles for face-on (0 deg) and edge-on (90 deg) galaxies. The dashed line at $W_{80, \mathrm{out}}/W_{80, \mathrm{gal}}=1.2$ represents the prior criterion used by \cite{carniani24} to identify outflows. According to this criterion, galaxies positioned to the left of this threshold line would not be recognized as having detectable outflows.

Because low-mass galaxies at early epochs often lack clearly defined disk structures, differentiating between edge-on and face-on orientations can be challenging. Therefore, to highlight the impact of galaxy orientation on more structured systems, we additionally examine galaxies with stellar masses $M_\star>10^{8.5}\mathrm{M}_\odot$ in the bottom panels of Fig.~\ref{fig:orientation}. The right panels show the complementary cumulative distributions of $W_{80, \mathrm{out}}/W_{80, \mathrm{gal}}$ values for four different orientation angles.
In the case of the full sample of galaxies, approximately 70\% of face-on galaxies and around 55\% of edge-on galaxies meet the JADES detectability prior. Hence, the detectability does depend on the viewing angle, but the difference in occurrence rate is of the order of 10-15\%. For galaxies within the mass range probed by JADES and assuming random orientations of their galactic planes, we find that $66.4\%$ of priors exhibit a $W_{80}$ ratio $>1.2$, thereby fulfilling the condition for outflow detectability.

When examining the sample of more massive galaxies below, we observe that orientation plays a significantly larger role. For these, disc-shaped galaxies, outflows viewed face-on are detectable in about 65\% of cases, whereas detectability drastically decreases to approximately 25\% for an edge-on orientation due to overlapping velocity distributions with the galactic gas.
While these galaxies are more massive than most of those observed in the JADES sample, it is important to note that TNG50 galaxies are generally more diffuse than expected at high redshift. Consequently, the angle dependence identified in the more massive simulated systems may well extend to real galaxies of lower mass.

\begin{figure*}
    \centering
    \includegraphics[width=0.48\linewidth]{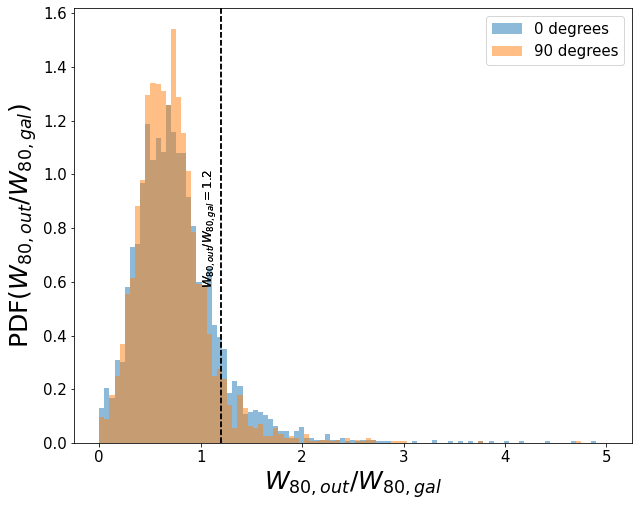}
    \includegraphics[width=0.48\linewidth]{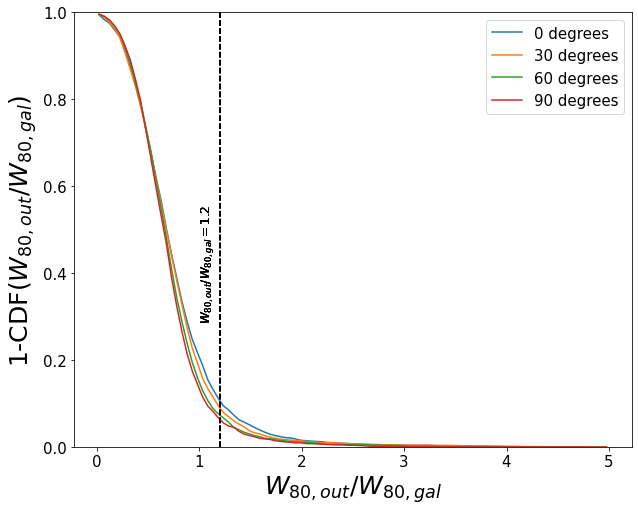}
    \caption{Left: Distribution of the ratios of velocity distribution widths between the outflows ($W_{80, \mathrm{out}}$) and the galaxy ($W_{80, \mathrm{gal}}$) in SERRA galaxies for a face-on orientation (0$^\circ$) and an edge-on orientation (90$^\circ$). The ratio of 1.2, indicated by the vertical line, represents minimal ratio used by JADES to separate outflows from galactic gas. \\
Right: Complement cumulative distribution of the width ratios for various galactic orientation angles.}
\label{fig:orientation_serra}
\end{figure*}

Finally, in fig.\ref{fig:orientation_serra}, we examine the orientation dependence of galactic outflow detectability in SERRA. This figure is analogous to the top row of fig.\ref{fig:orientation}. We notice that only about 10–15\% of SERRA outflows would be detectable, which is a factor of 3-4 lower than what is observed in the JADES study. Given that most SERRA galaxies have cooled into disc-like structures, one might expect a larger angular dependence compared to TNG50. Surprisingly, the opposite appears to be true: the difference in outflow detectability between face-on and edge-on orientations is only marginal.
For lower-mass galaxies, this unexpected result can be attributed to the high frequency of mergers, which dominate the outflow signatures. In a merger, the orientation of the tidal tail of the merger is independent of the galaxy orientation, thus resulting in a weak angular dependence of the identified outflows.
Interestingly, this lack of angular dependence persists even when the sample is restricted to only high-mass galaxies, as was done for the TNG50 sample in the bottom row of fig.~\ref{fig:orientation}\footnote{Given that restricting the sample to high-mass galaxies does not significantly alter the outcome, we have not plotted this subset separately.}. In this regime, major mergers are much less common, and yet no strong orientation dependence emerges. This likely reflects the longer merger timescales, which leave prominent merger remnants that obscure the underlying galactic morphology and, with it, any preferred outflow direction. An example is the massive SERRA galaxy shown in Fig.~\ref{fig:example_galaxies}, where residual rotational features appear misaligned with the galactic disc. This feature is present in a substantial fraction of SERRA galaxies and offers a natural explanation for the lack of orientation dependence. The origin of these frequently observed secondary discs is likely tied to merger events. Typically, SERRA galaxies are not fully disrupted and reassembled during mergers \citep{kohandel:2020, kohandel:2024}, instead, they tend to interpenetrate and gradually coalesce, allowing remnants of the merging systems—such as misaligned gaseous discs—to persist temporarily as distinct rotational components. Although some systems do exhibit clear directional outflows \citep{kohandel:2025}, such cases seem rare and do not appear to dominate the overall statistics. 

\subsection{Evolution of outflows over time \label{time-evolution}}

\begin{figure*}
    \centering
    \includegraphics[width=0.43\linewidth]{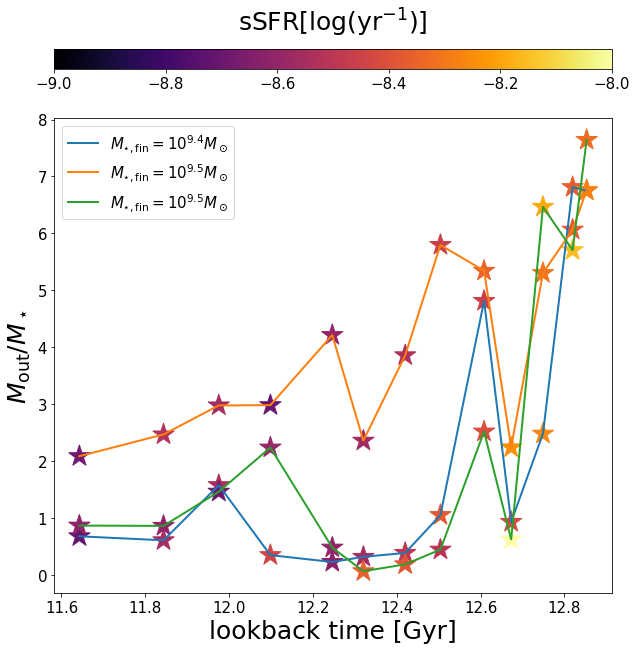}
    \includegraphics[width=0.48\linewidth]{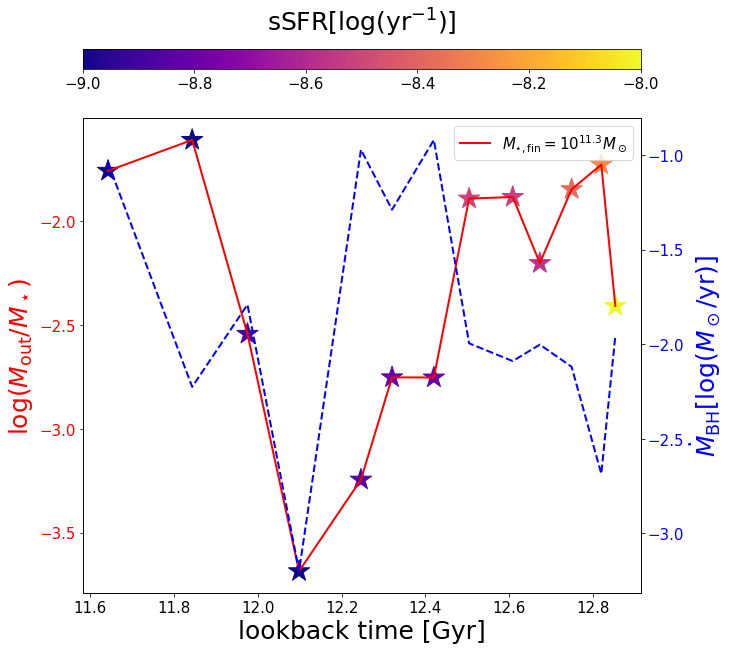}
    \caption{Left: Evolution of the ratio of outflow gas mass to galactic stellar mass for three medium sized galaxies as a function of lookback time with the sSFR at a given snapshot depicted on the color scale. Hereby $M_{\star,\mathrm{fin}}$ represents the stellar mass at the most recent time \\ Right: Evolution of the same ratio but on a logarithmic scale for one massive galaxy with the blue line representing the rate of black hole.}
    \label{fig:time_evolution}
\end{figure*}

In Fig.~\ref{fig:time_evolution}, we examine the temporal evolution of outflows within individual galaxies from the TNG50 simulation. To this end, we follow the most massive progenitors of selected galaxies across the redshift range considered, computing their outflow properties at each snapshot. Because our algorithm tracks galaxies that consistently remain the most massive within their host halos, we focus on a small set of representative cases for which a progenitor can be reliably identified at every snapshot.

In the left-hand plot, we highlight three galaxies that reach a final stellar mass of approximately $M_{\star, \mathrm{fin}} = 10^{9.5}\mathrm{M}_\odot$ at redshift $z=3$. This choice was made to ensure that the sample contains galaxies without an active AGN while being massive enough to have a history spanning multiple redshifts. From this group, we selected three examples that exhibited particularly strong evolution in their specific SFR (sSFR). We find that long-term increases in the relative mass of outflows correlate closely with increases in sSFR. However, there are notable abrupt increases in outflow activity that cannot be solely explained by star formation bursts; these are likely driven in part by tidal interactions with neighboring galaxies. The selected examples show a general trend of declining star formation and outflow activity at later times.

On the right-hand side, we show the evolution of outflows in one of the most massive galaxies in the TNG50 simulation, with a final stellar mass of $M_{\star, \mathrm{fin}} = 10^{11.5}\mathrm{M}_\odot$ at $z=3$. This massive galaxy exhibits notably different behaviour compared to less massive examples. Initially, we observe a similar decrease in sSFR and corresponding decline in outflow activity. However, outflows increase again at a lookback time of approximately $12.2$ Gyr (note the logarithmic scale, required due to large fluctuations). As indicated by the blue line representing the black hole growth rate, the resurgence of outflow activity at later times is likely driven by increased black hole accretion, resulting in powerful outflow jets from the central AGN.

In summary, low-mass galaxies, where outflows are predominantly supernova-driven, exhibit a close correlation between outflow mass and sSFR, with outflows declining at lower redshifts once the galaxy becomes more quenched. In massive galaxies, the same trend is observed initially; however, at later times, the central black hole emerges as the dominant driver of outflows. As a result, the outflow mass continues to rise toward lower redshifts, even as the sSFR declines substantially.

\section{Conclusions and Outlook \label{sec:Conclusions}}

Recent advances in observations opened new possibilities for examining galactic outflows in unprecedented detail.
In order to accurately interpret these observations, a comparison with their theoretical counterpart is crucial. For this purpose, we developed a novel methodology to study outflows in the immediate vicinity of galaxies drawn from two distinct simulations, TNG50 \citep{pillepich18b} and SERRA \citep{pallottini22}.
To extract these outflows, we employed a Gaussian mixture model to assign gas particles in and around each galaxy to different modes based on their properties, including spatial position, velocity, and star formation rate.
This approach allowed us to disentangle overlapping gas components and robustly separate outflowing from galactic gas.
We have then estimated the outflows properties and compared with those inferred from the survey JADES \citep{carniani24}. We obtained the following key results:

\begin{itemize}

\item  The outflow masses estimated in both the TNG50 and SERRA simulations span a wide range, from $10^5$ to $10^9$~$\mathrm{M}_\odot$, depending on galaxy mass. At the stellar masses of JADES galaxies ($\sim 10^8~\mathrm{M}_\odot$), the average outflow mass is $10^{7.5}~\mathrm{M}_\odot$  and $10^7~\mathrm{M}_\odot$ for TNG50 and SERRA, respectively, and they are about 0.5–1 dex higher than the median values inferred from observations. This offset may suggest that optical lines trace only a fraction of the total outflow. However, considering the full distribution of outflow masses in the simulations, we note that the discrepancy is not severe and predictions and observations agree within $1$–$2\sigma$. We also note that the JADES sample by \citealt{carniani24} includes only 14 galaxies, which is not sufficient to obtain a complete distribution of outflow mass estimates.

\item Outflow velocities inferred from simulations are, on average, a factor of about $10$ lower than those observed. Notably, a subset of targets shows velocities that exceed the predictions of both SERRA and TNG50. This discrepancy likely reflects model assumptions and parameter choices that systematically underpredict wind velocities, such as the halo mass dependent wind speed in TNG50.

\item The observed metallicities of outflows tend to be lower than those of their host galaxies, suggesting that estimates of outflow mass derived from metal-line emission (such as [OIII]) could be systematically underestimated, potentially by a factor of two to eight.

\item In TNG50, the detection of outflows as a broad spectral component is strongly affected by galaxy orientation. Face-on systems generally exhibit higher detectability than edge-on systems. While on average the difference of detectability is $15\%$ it increases to $40\%$ for more massive ($M_\star \gtrsim 10^{8.5}\mathrm{M}_\odot$) galaxies, which tend to have a disc-like morphology. In contrast, the more merger-driven outflows in SERRA do not display a comparable orientation dependence, suggesting that their kinematic signatures are dominated by merger-induced dynamics.

\item For intermediate-mass galaxies ($M_\star \approx 10^{9.5}\mathrm{M}_\odot$), we observe a decrease in outflow activity from $z=6$ to $z=3$, likely associated with a decreasing star formation activity, as evidenced by declining sSFR. In contrast, massive galaxies ($M_\star \approx 10^{11}\mathrm{M}_\odot$) show a resurgence of outflow activity at later times due to increased feedback from central AGN.

\end{itemize}

This study has compared outflows in the immediate vicinity of galaxies with their observational counterparts from JWST. By employing two independent simulations, we were able to disentangle intrinsic physical trends from model-specific features, thereby increasing the robustness of our interpretation.

\section*{Data availability}

Data directly related to this publication and its figures is available on request from the corresponding author. The IllustrisTNG simulations, including TNG50, are publicly available and accessible at \url{www.tng-project.org/data} \citep[see][]{nelson19a}. 

\begin{acknowledgements}
We acknowledge support by the European Union's HE ERC Starting Grant No. 101040227 - WINGS (PI: Carniani).
Any dissemination of results must indicate that it reflects only the author's view and that the Commission is not responsible for any use that may be made of the information it contains.
We acknowledge the CINECA award under the ISCRA initiative, for the availability of high-performance computing resources and support from the Class B project SERRA HP10BPUZ8F (PI: A. Pallottini).
We gratefully acknowledge the computational resources of the Center for High Performance Computing (CHPC) at SNS.
The authors extend their gratitude to Ruediger Pakmor for insightful discussions that have contributed to the development of this work, in particular, on the characteristics of the TNG50 dataset.
\end{acknowledgements}

\bibliographystyle{aa} 
\bibliography{refs}
\begin{appendix}
\section{Verifying the Gaussian outflow selection \label{app:algorithm}}
To evaluate the performance of our outflow–selection model, we applied it to a synthetic galaxy system for which the ground truth is known. 
We used a simplified model in which the gas in the galaxy is represented by $\sim10{^6}$ point-like synthetic emitting sources (hereafter referred to as particles) distributed according to a prescribed geometric configuration. For the disk component, we assumed that the particles are distributed in a thin disk following an exponential surface density profile. We adopted a scale radius of 1.5 kpc and assumed a disk height equal to one fifth of the scale radius. Each particle in the disk is assigned a three-dimensional velocity vector composed of the expected circular velocity for an exponential disk with a dynamical mass of ${\rm 10^{10}~\mathrm{M}_\odot}$ (\cite{Binney2008} ; see also eq.~6\citealt{Parlanti2023} ) and a random velocity component of $\sim50~{\rm km~s^{-1}}$.  We adopted a bi-conical geometry for the outflow, with an opening angle of 45 degree and an extent of 5 kpc. The particles were distributed throughout the volume of the two cones under the assumption that the average volume density of the outflowing gas decreases outward with the square of the radius. This is consistent with a scenario in which the mass outflow rate is constant with time \citep{Lutz2020}. The outflow particles were also assigned a 3D  velocity vector composed of a constant radial velocity of $500~{\rm km~s^{-1}}$ and an additional random 3D velocity component with a magnitude of $300~{\rm km~s^{-1}}$.

\begin{figure}
\includegraphics[width=0.49\textwidth]{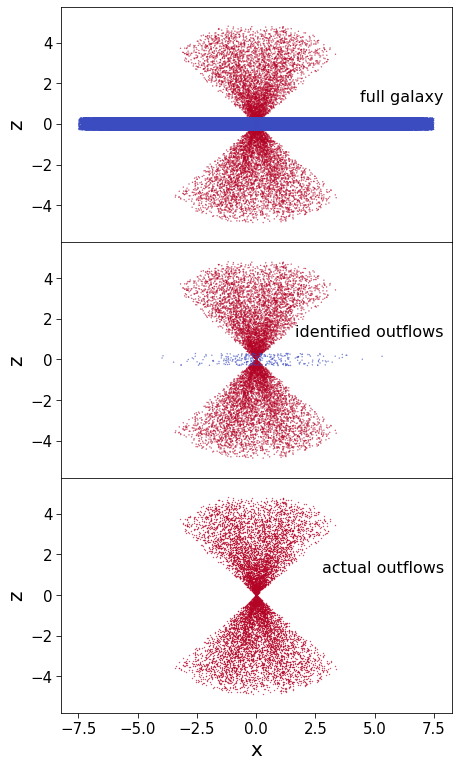}
\caption{Test of the outflow selection algorithm. Top: Galaxy particles (blue) together with outflow particle (red), centre: particles identifies as outflows using the Gaussian mixture model approach, bottom: ground truth, i.e. actual outflows.}
    \label{fig:selection_test}
\end{figure}

The model galaxy used for testing is shown in Fig.~\ref{fig:selection_test}. It contains $10^6$ particles representing the galactic gas and $10^4$ particles associated with the outflow. We apply our algorithm, as described in Section~\ref{subsec:identfication}, to the full gas sample. In this setup, however, we modify the definition of the galactic gas: instead of identifying it as the mode with the smallest average distances from the centre, we define it as the mode comprising particles with the lowest average flow velocities. This adjustment reflects that, in our example galaxy, the galaxy is more spatially extended than the outflow regions.

We find that the algorithm performs robustly (see central panel of Fig.~\ref{fig:selection_test}), despite the large dynamic range, spanning two orders of magnitude—between galactic and outflow particles. In total, 308 particles, corresponding to $3.1\%$ of the outflow sample, are falsely identified as galactic gas. These are typically located close to the overlapping regions between the galaxy and its outflows. Conversely, only 61 particles ($0.6\%$) of the outflow population are missed, i.e. incorrectly classified as galactic. Overall, the algorithm demonstrates excellent performance in distinguishing outflows from galactic gas, even in regions such as the galactic centre where both components are strongly interwoven. 

\section{Contribution of TNG50 wind particles to galactic outflows \label{app:wind}}
As described in Sec.~\ref{subsec:tng50}, outflows from star-forming regions in the TNG50 simulation are initialized as non-interacting wind particles that propagate through the galaxy before they hydrodynamically recouple once the density falls below $0.005\mathrm{cm}^{-3}$. This means that we miss some of the outflows that have not recoupled yet if we consider only gas particles. The goal of this appendix is to investigate the contribution of wind particles to the total outflow.

\begin{figure*}
    \centering    \includegraphics[width=0.49\linewidth]{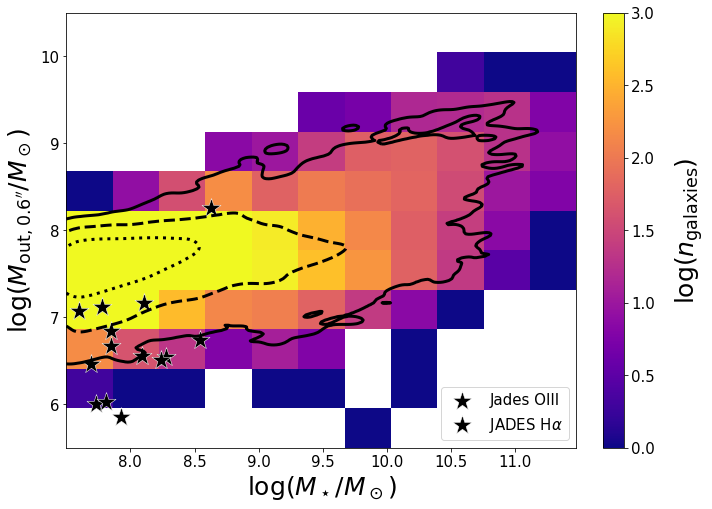}
    \includegraphics[width=0.49\linewidth]{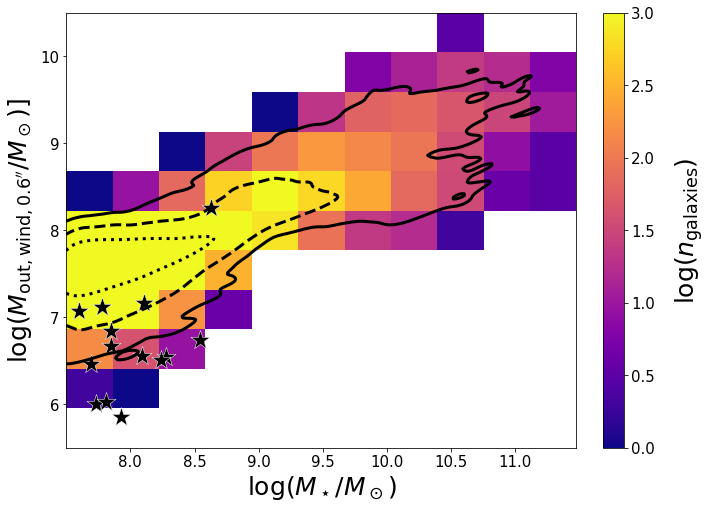}
    \caption{Distribution of the TNG50 outflow mass as a function of galactic stellar mass, without wind particles (left) and with wind particles included (right).}
    \label{fig:wind_mass}
\end{figure*}

In Fig.~\ref{fig:wind_mass} we compare the outflow mass for galactic outflows with and without wind-particles as a function of galactic stellar mass. The plot on the left only contains the mass of the gas particles such that the contours match those in Fig.~\ref{fig:Mout}, while the plot on the right also includes wind particles. As one can see the contribution of wind is negligible to the outflow-mass in the galactic stellar mass range $10^{7.5}-10^{8.5}\mathrm{M}_\odot$ encompassing the range of most galaxies observed in JADES. However, for larger galaxies, the contribution of wind particles becomes increasingly significant lifting up the lower limit of the outflow masses while leaving the upper limit mostly unchanged. This indicates that in the galaxies with higher outflows the wind-particles are typically already recoupled while very low outflow masses tend to reflect the fact that most of the outflows have not yet recoupled rather than a physically lower value. 

\begin{figure*}
    \centering
    \begin{subfigure}{0.33\linewidth}
        \centering
        \includegraphics[width=\linewidth]{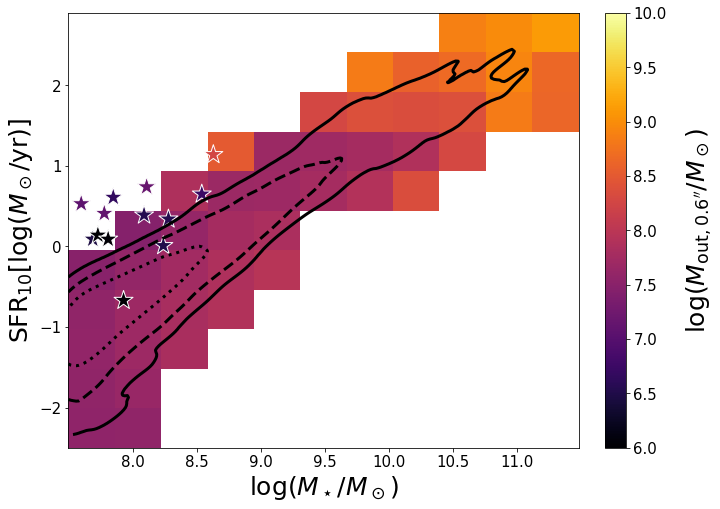}
        \subcaption{TNG50 outflows without wind}
    \end{subfigure}
    \begin{subfigure}{0.33\linewidth}
        \centering
        \includegraphics[width=\linewidth]{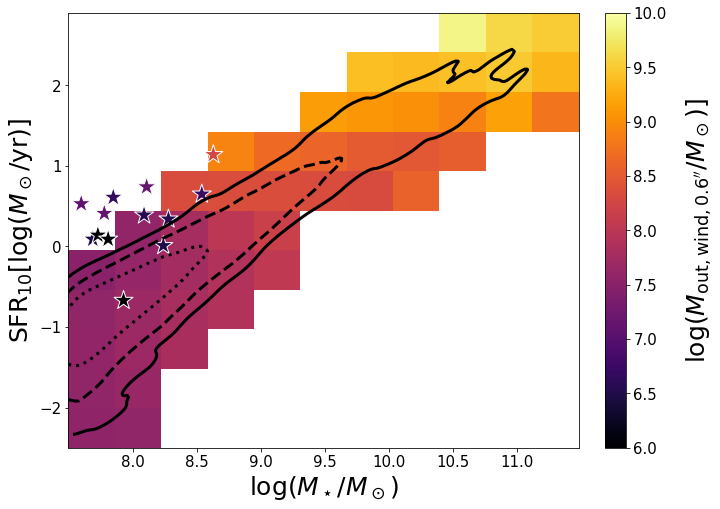}
        \subcaption{TNG50 outflows with wind}
    \end{subfigure}
    \begin{subfigure}{0.33\linewidth}
        \centering
        \includegraphics[width=\linewidth]{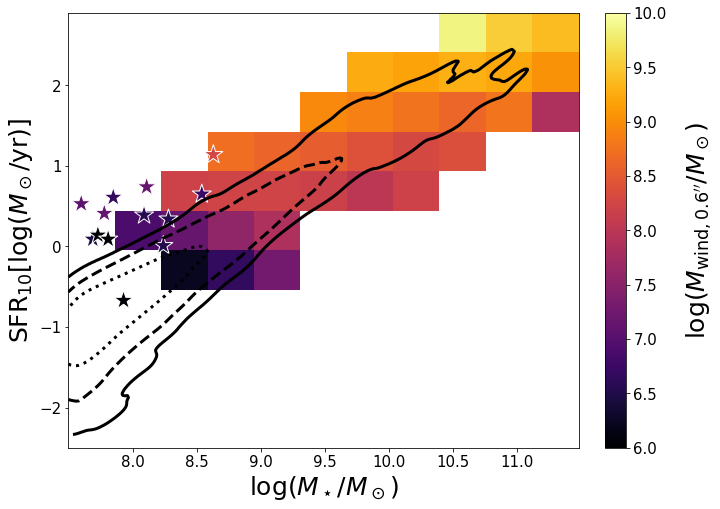}
        \subcaption{TNG50 outflows only wind}
    \end{subfigure}
    
    \caption{Median outflow mass as a function of galactic stellar mass and SFR without wind particles (left), with wind particles (center) and only including wind particles (right).}
    \label{fig:wind_sfr}
\end{figure*}

Given that wind particles contribute significantly to the outflow mass, particularly in more massive galaxies, in Fig.\ref{fig:wind_sfr} we re-examine the correlation between outflows and SFR previously shown in Fig.~\ref{fig:SFR}, this time including wind particles. For galaxies with $M_\star \gtrsim 10^9\mathrm{M}_\odot$, the correlation with SFR indeed becomes more apparent, as indicated by a clear colour gradient at fixed stellar mass. This trend is even more pronounced in the right-hand panel, which shows only the wind particles. These findings support our earlier hypothesis that the apparent lack of correlation between outflow mass and SFR in higher-mass galaxies is primarily due to the most recent outflows, driven by star formation within the past 10Myr,having not yet hydrodynamically recoupled.
\end{appendix}
\end{document}